\documentclass[twocolumn]{aastex631}

\newcommand\name{{\it Eos}}
\newcommand\namesp{{\it Eos}\ }

\shorttitle{Milky Way's eccentric constituents}
\shortauthors{Myeong et al.}
\graphicspath{{./}{figures/}}

\begin{document}

\title[Milky Way's eccentric constituents]{Milky Way's eccentric constituents with \textit{Gaia}, APOGEE \& GALAH}

\correspondingauthor{GyuChul Myeong}
\email{gyuchul.myeong@cfa.harvard.edu}


\author[0000-0002-5629-8876]{G.~C. Myeong}
\affiliation{Harvard-Smithsonian Center for Astrophysics, 
60 Garden Street, 
Cambridge, MA 02138, USA}

\author[0000-0002-0038-9584]{Vasily Belokurov}
\affiliation{Institute of Astronomy, University of Cambridge, 
Madingley Road, 
Cambridge, CB3 0HA, UK}
\affiliation{Center for Computational Astrophysics, Flatiron Institute, 
162 5th Avenue, 
New York, NY 10010, USA}

\author[0000-0001-5200-3973]{David~S. Aguado}
\affiliation{Dipartimento di Fisica e Astrofisica, Univerisità degli Studi di Firenze, 
via G. Sansone 1, 
I-50019 Sesto Fiorentino, Italy}
\affiliation{INAF/Osservatorio Astrofisico di Arcetri, 
Largo E. Fermi 5, 
I-50125 Firenze, Italy}

\author[0000-0002-5981-7360]{N.~Wyn Evans}
\affiliation{Institute of Astronomy, University of Cambridge, 
Madingley Road, 
Cambridge, CB3 0HA, UK}

\author[0000-0003-2352-3202]{Nelson Caldwell}
\affiliation{Harvard-Smithsonian Center for Astrophysics, 
60 Garden Street, 
Cambridge, MA 02138, USA}

\author{James Bradley}
\affiliation{Institute of Astronomy, University of Cambridge, 
Madingley Road, 
Cambridge, CB3 0HA, UK}




\begin{abstract}

We report the results of an unsupervised decomposition of the local stellar halo in the chemo-dynamical space spanned by the abundance measurements from APOGEE DR17 and GALAH DR3. In our Gaussian Mixture Model, only four independent components dominate the halo in the Solar neighborhood, three previously known \textit{Aurora}, \textit{Splash} and Gaia-Sausage/Enceladus (GS/E) and one new, \name. Only one of these four is of accreted origin, namely the GS/E, thus supporting the earlier claims that the GS/E is the main progenitor of the Galactic stellar halo. We show that \textit{Aurora} is entirely consistent with the chemical properties of the so-called Heracles merger. In our analysis in which no predefined chemical selection cuts are applied, \textit{Aurora} spans a wide range of [Al/Fe] with a metallicity correlation indicative of a fast chemical enrichment in a massive galaxy, the young Milky Way. The new halo component dubbed \namesp is classified as \textit{in situ} given its high mean [Al/Fe]. \namesp shows strong evolution as a function of [Fe/H], where it changes from being the closest to GS/E at its lowest [Fe/H] to being indistinguishable from the Galactic low-$\alpha$ population at its highest [Fe/H]. We surmise that at least some of the outer thin disk of the Galaxy started its evolution in the gas polluted by the GS/E, and \namesp is an evidence of this process.

\end{abstract}



\section{Introduction} \label{sec:intro}

One curious thing to ponder about our Galaxy's continuous evolution is that the guises the Milky Way takes as it transforms all remain in place -- albeit somewhat hidden -- for us to make sense of.
Today, with new data that let us see the Galaxy in phase space (thanks to the ESA's space observatory {\it Gaia}~\citep{GaiaEDR32021}) and in chemical space (thanks to massive high-resolution surveys like Apache Point Observatory Galactic Evolution Experiment~\citep[APOGEE;][]{Abdurrouf_APOGEE2022} and Galactic Archaeology with HERMES~\citep[GALAH;][]{Buder_GALAH2021}), it is possible to reveal the Milky Way's forgotten states and discarded guises. The Galactic halo specifically has remained the centre of intense attention thanks to its capacity to keep orbital information neatly preserved for every star deposited -- this at a clear advantage compared to the studies of the disk where the past state is less straightforward to decipher due to the effects of radial mixing~\citep[see e.g.,][]{Sellwood2002,Schonrich2009,Loebman2011,Minchev2013,Frankel2020}.  

The halo is vast and even though its deep trawl only just started, the haul offered up by \textit{Gaia} has been copious. For example, a large number of low-mass accreted systems were detected in the Solar neighbourhood~\citep[see e.g.,][]{Myeong2018clumps,Myeong2018_OmC,Koppelman2019,Ohare2020,Borsato2020,Limberg2021,Lovdal2022} and beyond~\citep[see e.g.,][]{Malhan2018ghostly,Ibata2019abyss,Palau2019,Naidu2020entirely,Yuan2020,Ibata2021charting}. On the other hand, clear evidence for remnants of more massive galaxies consumed by the Milky Way appears rather scarce indicating a modest and restrained accretion course our Galaxy has followed. 

So far, only one ancient and massive intruder has been identified unambiguously, with its body now nothing but an enormous cloud of tidal wreckage scattered throughout the inner Milky Way. The existence of a vast debris structure left behind by this merger event, known today as the Gaia Sausage/Enceladus (GS/E), was suggested before {\it Gaia}~\citep[see][for the historical development]{Ev20}.
In particular, \citet{Deason2013} argued that the rapid transition in the Galactic stellar halo structural properties at break radius of 20-30 kpc is likely associated with the apocentric pile-up of a relatively early (8-10 Gyr ago) and single massive accretion event. It is hypothesised that the progenitor dwarf galaxy was massive enough for its orbit to shrink and radialise quickly as a result of a complex interplay between dynamical deceleration, host recoiling and self-friction~\citep[][]{Vasiliev2022radialization}. Sinking deep in the heart of the Milky Way, the dwarf sprayed the bulk of its stars in a region enclosed by its last apo-centre, i.e. some $\sim30$\,kpc~\citep[see][]{Deason2018pileup}. As a result, the region of the Galactic halo within this so-called break radius~\citep[see][]{Deason2011} is inundated with the GS/E stars, consistently showing up as the most striking sub-structure even in relatively small volumes around the Sun~\citep[see examples of pre-{\it Gaia} hints in e.g.,][]{Chiba1998,Brook2003,Meza2005,Nissen2010,Hawkins2015}. 
Subsequently, the {\it Gaia} data made crystal clear the unusually strong radial anisotropy of the relatively metal-rich GS/E debris and helped to reveal its dominance in the Solar neighbourhood~\citep[see][]{Belokurov2018,Myeong2018_actionHalo,Haywood2018}. The genesis of the GS/E debris cloud was also unambiguously confirmed by its unique chemical fingerprint~\citep[see e.g.,][]{Helmi2018,Mackereth2019} and the large group of globular clusters (GCs) associated with it~\citep[see e.g.,][]{Myeong2018_SauGC,Myeong2019_Sq,Massari2019}. Close to the Sun, the GS/E structure has been thoroughly scrutinised both kinematically and chemically~\citep[see][]{Necib2019,Evans2019,Sahlholdt2019,Das2020,Monty2020,Kordopatis2020,Feuillet2020,Molaro2020,Carollo2021,Aguado2021,Matsuno2021,Bonifacio2021,Feuillet2021,Buder2021}. Outside of the Solar neighborhood, fewer studies exist, nonetheless the global structure of the GS/E cloud is starting to come into focus~\citep[see e.g.,][]{Simion2019,Iorio2019,Lancaster2019,Naidu2020entirely,Bird2021,Balbinot2021,Iorio2021,Wu2022}.

Two lines of enquiry help to cement the unique place the GS/E merger holds in the history of our Galaxy: its mass and the time of its arrival. Using established mass-metallicity relations~\citep[see][]{Kirby2013}, the total stellar mass of the progenitor dwarf before disruption can be gleaned from the typical metallicity of its stellar population~\citep[][]{Belokurov2018,Feuillet2019,Naidu2021}. An independent mass constraint can be obtained by scaling up the total number of globular clusters associated with the merger~\citep[][]{Myeong2018_SauGC,Myeong2019_Sq,Massari2019,Forbes2020,Callingham2022}. A third method of constraining the total stellar mass is via the analysis of the GS/E chemical pattern~\citep[e.g.,][]{Fernandez2018,Helmi2018,Vincenzo2019,Mackereth2019}. All of the above estimates agree that the stellar mass of the GS/E progenitor had to be of order of few\,$\times10^8 M_{\odot}$. This can be contrasted with the total stellar halo mass measurements of $\sim10^9 M_{\odot}$~\citep[see][]{Deason2019halomass,Mackereth2020halomass}. As the comparison shows, the GS/E merger must have contributed a significant fraction of the total stellar halo, perhaps as much as a half (for the inner $\sim50$\,kpc region) in agreement with local~\citep[][]{Belokurov2018} and global~\citep[][]{Iorio2019} estimates. In other words, a simple budget of the Galactic stellar halo leaves little room for other significant accretion events. 

The dominance of the GS/E merger also makes sense when the timeline of the Galactic accretion is considered. Constraining the time of the accretion is far from straightforward. Not many asteroseismological ages exist for relatively metal-poor old stars~\citep[see][]{Montalban2021,Grunblat2021,Alencastro2022}. These direct, albeit scarce, measurements place the epoch of the GS/E merger at 8-10 Gyr ago, in agreement with White Dwarf cooling ages~\citep[][]{Fantin2021}. The age estimates for a bulk of GS/E stars based on main-sequence-turn-off stars~\citep{Bonaca2020} and red giants~\citep{Das2020} also show good agreements with this time frame. Alternatively, the time of accretion can be obtained via the inference of the star-formation history (SFH) of the progenitor galaxy~\citep[e.g.,][]{Helmi2018,Vincenzo2019,Mackereth2019,Sanders2021}. These modelling attempts agree that the dwarf's SFH was shut down between 8 and 10 Gyr ago. Finally both the progenitor mass and the accretion time can be gauged by comparing the observed redshift $z=0$ properties of the tidal debris cloud with its numerical counterparts in galaxy formation simulations~\citep[see][]{Belokurov2018,Fattahi2019,Bignone2019,Elias2020,Dillamore2022}. Across different simulation suites, these studies find that the accretion events that lead to creation of a highly radially anisotropic metal-rich stellar halo structure near the Sun typically take place 7-10 Gyr ago and that the progenitor galaxies bring between $10^8$ and $10^9 M_{\odot}$ of stars.

Does this imply that the GS/E merger is the only significant merger our Galaxy has experienced? Not necessarily. Remnants of even older interactions could potentially be hidden in the inner Galaxy, i.e. the area known as the {\it bulge} and, indeed, several such claims have been made recently~\citep[e.g.,][]{Kruijssen2019,Forbes2020,Horta2021}. Additionally, more recent or ongoing accretion events may be lurking in the Milky Way's outskirts. There are at least two such relatively recent and relatively massive accretion events: that of the Sagittarius dwarf galaxy~\citep[see][]{Ibata1994,Majewski2003} and of the Magellanic System~\citep[see][]{Wannier1972,Mathewson1974}. The GS/E, the Magellanic Clouds and the Sgr dwarf are all estimated to have been hosted in dark matter halos with masses not far off $10^{11} M_{\odot}$~\citep[see e.g.,][]{Fattahi2019,Erkal2019,Gibbons2017}. In addition, tidal debris from several systems with at least an order of magnitude lower mass have also been found, e.g., Sequoia~\citep[see][]{Myeong2018_OmC,Myeong2019_Sq,Matsuno2019}.

The analysis reported here is inspired by the recent discoveries of a large number of diverse halo sub-structures and the need for a systematic tally of past mergers with a focus on the most important events. To our knowledge, not many attempts have been made to build a well-defined census of Galactic accretion remnants. Taking advantage of a rich but relatively small Galactic Globular Cluster dataset, several groups sought to apportion the known GCs to a small number of significant mergers~\citep[][]{Massari2019,Kruijssen2019,Forbes2020,Callingham2022}. 
However, the choices as to the decision boundaries and the number of independent components were made by hand, and therefore, unfortunately,  were neither objective or reproducible (in e.g., a numerical simulation). 
In a similar vein, \citet{Naidu2020entirely} suggested a break-down of the entire stellar halo accessible to the H3 survey into individual accretion events 
but their substructure membership, similarly to the works above, was designed by hand.
Most similar to our study are the works of \citet{Das2020} and \citet{Buder2021} who both aimed to extract the footprint of the local GS/E tidal debris in the space spanned by chemical abundances. They proceeded by modelling high-quality APOGEE and GALAH samples with a mixture of Gaussians and identifying the component representing the GS/E against the background of other Galactic populations.

The aim of our study is to build an unbiased and self-consistent inventory of the stellar halo in the vicinity of the Sun. Our main objective is to identify and characterise all of the significant independent halo components without postulating their number and their behaviour a priori. For the task in hand we will also use an approach based on a mixture of Gaussians. The stars belonging to the \textit{in situ} populations such as the disk are expected to dominate the halo denizens by almost two orders of magnitude. To avoid wasting model components on known Galactic populations at the expense of the objects of our experiment, we limit the scope of our study solely to stars on orbits with high eccentricity. Thus, the stellar sample considered consists only of stars unambiguously attributed to the Galactic halo. While demanding high eccentricity ensures that the halo sample is pure, it necessarily leads to incompleteness because tidal debris with significant angular momentum will be excised. In terms of the accreted sub-structures, there are plenty of local examples with both prograde~\citep[e.g., Helmi and the associated S2 stream,][]{Helmi1999,Myeong2018clumps,Aguado21a} or retrograde~\citep[e.g., Sequoia,][]{Myeong2018_OmC,Myeong2019_Sq,Matsuno2019} orbital motion. 

Note that thanks to {\it Gaia} we are now certain that stars on high-eccentricity, or halo-like, orbits do not necessarily have to be accreted. As first demonstrated by \citet{Bonaca2017} roughly half of the local stellar halo is of high metallicity, i.e. [Fe/H]$>-1$\,dex and is likely of \textit{in situ} origin. These \textit{in situ} halo stars had of course been seen before, e.g., as the high-$\alpha$ halo sequence by \citet{Nissen2010} and \citet{Hawkins2015} as later confirmed by e.g., \citet{Haywood2018}. The origin of the \textit{in situ} halo stars with [Fe/H]$>-1$\,dex has been linked to the GS/E accretion event~\citep[][]{Gallart2019,Dimatteo2019,Belokurov2020}. These stars were born in the ancient high-$\alpha$ disk of the Milky Way before the merger and subsequently ``splashed'' out of their circular orbits during the interaction with the GS/E progenitor, a scenario ubiquitous in numerical simulations of galaxy formation~\citep[see][]{Belokurov2020,Grand2020,Dillamore2022}. Not all of the \textit{in situ} halo is splashed high-$\alpha$ disk. As \citet{Belokurov2022} demonstrate using APOGEE DR17 data, the Milky Way stars with [Fe/H]$<-1.5$\,dex were likely formed in a hot and messy pre-disk state that later phase-mixed to form a centrally concentrated spheroidal \textit{in situ} halo component dubbed \textit{Aurora}. \citet{Conroy2022} also see evidence for an ancient Galactic pre-disk population in the low-metallicity sample of H3 stars.

This paper is structured as follows. Section~\ref{sec:data} describes the two sets of selection cuts applied to the APOGEE and the GALAH data to produce the samples of high-quality abundance measurements analysed. As the APOGEE DR17 reports abundances with noticeably smaller uncertainties compared to GALAH we consider the APOGEE sample as our primary dataset. Section~\ref{sec:method} briefly gives the details of the Gaussian Mixture Model methodology used here. We analyse the components extracted from APOGEE and GALAH in Section~\ref{sec:results}. Our results are placed in context and summarised in Section~\ref{sec:summ}. 

\section{Data} \label{sec:data}

We use the main and Value-Added-Catalogue (VAC) data of APOGEE Data Release (DR) 17~\citep{Abdurrouf_APOGEE2022} which provides stellar parameters, radial velocity and abundances for up to 20 chemical species for 372,458 unique targets (main red star sample). 
It is cross-matched with \textit{Gaia} EDR3~\citep{GaiaEDR32021}. Using distance estimates from \citet{Bailer-Jones2021}, orbital dynamics are calculated using \textsc{AGAMA}~\citep{Vasiliev_AGAMA2019} with the best fitting axisymmetric potential of \citet{McMillan2017}.

The sample is cleaned by rejecting any stars with {\tt\string STAR\_BAD} flag in {\tt\string ASPCAPFLAG}, and {\tt\string PROGRAMNAME $=$ magclouds}. Only the main red star samples with {\tt\string EXTRATARG $= 0$} are used. To obtain a red giant star sample, {\tt\string logg} $< 3.0$ is adopted. It is further filtered with {\tt\string x\_fe\_flag} $= 0$, {\tt\string x\_fe\_err} $< 0.1$\,dex for almost all elements {\tt\string x} used in our study. The exception is  [Ce/Fe], for which {\tt\string ce\_fe\_err} $< 0.15$\,dex is used instead of $0.1$. 
To focus on the halo or halo-like Galactic components such as the GS/E~\citep[][]{Belokurov2018,Haywood2018,Myeong2018_actionHalo,Myeong2018_SauGC} and the \textit{Splash}~\citep{Belokurov2020}, an eccentricity $> 0.85$ cut is adopted to obtain stars on nearly radial orbits~\citep[see e.g.,][]{Belokurov2018,Myeong2018_actionHalo,Myeong2019_Sq,Mackereth2019,Buder2021}. We also require the orbital apocenter $> 5$\,kpc to minimize the contamination from the bulge populations~\citep[see e.g.,][]{Horta2021,Buder2021}. 
In addition, the distance error $< 1.5$\,kpc, and orbital energy  $E < 0$\,km$^{2}$\,s$^{-2}$ cuts are applied. 

For our APOGEE-\textit{Gaia} sample, we use [Fe/H], [$\alpha$/Fe], [Al/Fe], [Ce/Fe], [Mg/Mn], and the orbital energy ($E$) information, forming a 6-dimensional chemo-dynamical space to study with Gaussian Mixture Modelling (GMM). There are a total of $\approx$1700 APOGEE stars that pass our quality cuts\footnote{We note that this is $\approx1.3\%$ of all APOGEE stars that pass the same quality cut but has orbital eccentricity $<0.85$}. The choice of our 6-dimensional space is motivated by the desire to include dynamical properties ($E$) as well as the ``most reliable'' (or ``reliable'' if necessary) elemental production channels for the categories of odd-Z (Al), $\alpha$- ($\alpha$, Mg), iron-peak (Mn), and $s$-process (Ce) elements, following APOGEE's recommendation for giant stars.

As a cross-check, we also use the main and the VAC data of GALAH DR3~\citep{Buder_GALAH2021} which contain 678,423 spectra of 588,571 mostly nearby stars. The catalogues provide stellar parameters, radial velocity, and abundances for up to 30 chemical species. The VAC also provides various additional information including the distance estimates~\citep[see also][]{Sharma2018} and Galactic orbital dynamics calculated using \textsc{galpy}~\citep{Bovy2015_galpy} with the \citet{McMillan2017} potential. 

We adopt the recommendations from the GALAH Collaboration for the choice of data columns and quality cuts\footnote{\url{https://www.galah-survey.org/dr3/using_the_data/}}. Our sample is first filtered with {\tt\string snr\_c3\_iraf} $> 30$, {\tt\string flag\_sp} $= 0$. We adopt similar cuts from our APOGEE sample, with the exception of using {\tt\string e\_x\_fe} $< 0.2$\,dex for elements {\tt\string x} used for our study.

For our GALAH-\textit{Gaia} sample, we use [Fe/H], [$\alpha$/Fe], [Na/Fe], [Al/Fe], [Mn/Fe], [Y/Fe], [Ba/Fe], [Eu/Fe], [Mg/Cu], [Mg/Mn], [Ba/Eu], as well as energy $E$. This includes similar tracers as the APOGEE sample, plus the $r$-process element (Eu) available in GALAH. Therefore for GALAH, the GMM is applied to a 12-dimensional chemo-dynamical space sampled with a total of $\approx$1100 stars that passed the above quality cuts\footnote{$\approx1.6\%$ of all GALAH stars that pass the same quality cut but has orbital eccentricity $<0.85$}.

\section{Method} \label{sec:method}

Gaussian Mixture Modelling is an unsupervised probabilistic modelling technique. It assumes that the overall distribution of some (unlabelled) data is the sum of a finite number of Gaussian distributions with unknown parameters. The optimal (maximum likelihood) model parameters can be searched for using Expectation-Maximization (EM) iterations. For our modelling, we use the \textsc{Extreme Deconvolution}~\citep{Bovy2011_XD} (XD) algorithm. One attractive feature of XD is the ability to include uncertainties associated with the observed data. The maximum number of EM iterations for each XD run is $10^9$ by default

We run XD multiple times with different numbers of Gaussian components, from $N = 1$ to $10$. For each model with $N$ Gaussians, we repeat the fit 100 times with different initial parameters randomly drawn from the parameter space. This is to insure that the algorithm obtains the true maximum likelihood, not a local maximum. The whole session of 100 random runs is also repeated multiple times to verify convergence to the best-fitting case.
%
%
We check the results of the XD fitting with a different GMM package from \textsc{Scikit-learn}~\citep{scikit-learn} to demonstrate consistency.

The APOGEE sample provides abundances with noticeably smaller uncertainties as compared to GALAH. Therefore, the APOGEE-\textit{Gaia} data serves as the primary sample for our study. The difference in the quality of the data is apparent in the overall level of scatter seen in each dataset\footnote{For example, in panels a) of Figures~\ref{fig:apogeeplanes} and \ref{fig:galahplanes_1}. Note the difference in tightness of the well-known high- and low-$\alpha$ sequences in the two datasets.}. 

To avoid potential over-fitting, we compute the Bayesian Information Criterion~\citep{Schwarz1978_BIC} (BIC) and the Akaike Information Criterion~\citep{Akaike1974_AIC} (AIC) for each XD run. We use the BIC score to assess the necessary number of components for each model. The lowest BIC score (highest maximum log-likelihood among all XD runs) indicates the GMM with 7 components as the optimal choice. Although the BIC scores for 5 to 8 Gaussian components cases appear relatively comparable, the 7 Gaussians case always scores the lowest BIC consistently from our repeated trials of XD fitting sessions. Since we repeat the XD fitting 100 times during each trial (with randomized initial parameters) for each model with $N$ Gaussians, we check the overall distribution of the score in each case. When compared across all models (from $N = 1$ to $10$), the change in the BIC distribution shows that the median and the highest score are also the lowest at $N=7$ case in addition to the fact that it achieves the overall lowest (preferred) BIC value.

\begin{figure}
\begin{center}
{\includegraphics[width=90 mm, angle=0]{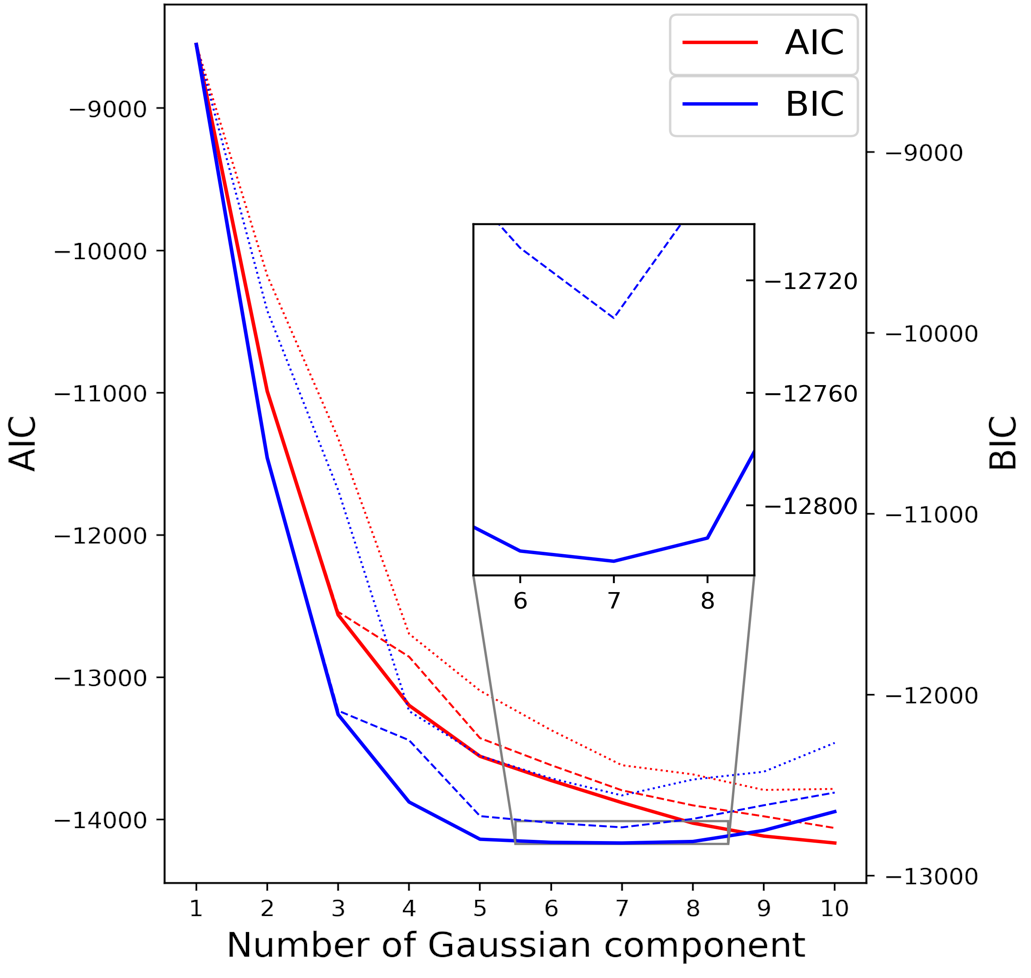}}
\end{center}
\caption{Red (blue) curves show AIC (BIC) performance for the APOGEE sample as a function of the number of Gaussian components. For each number of components, the GMM fit is performed a large number of times starting from random initial conditions to ensure convergence and to investigate scatter in obtained solutions. For each number of components probed, solid line gives the lowest (best) AIC/BIC score, dashed line corresponds to the median, and dotted to the highest score. The BIC test indicates the GMM with 7 Gaussian components as the optimal case. The zoom-in panel shows the BIC result around the best score.}
\label{fig:aicbic}
\end{figure}

\begin{figure*}
\begin{center}
{\includegraphics[width=185mm,angle=0,trim={ .cm .cm .cm 0cm},clip]{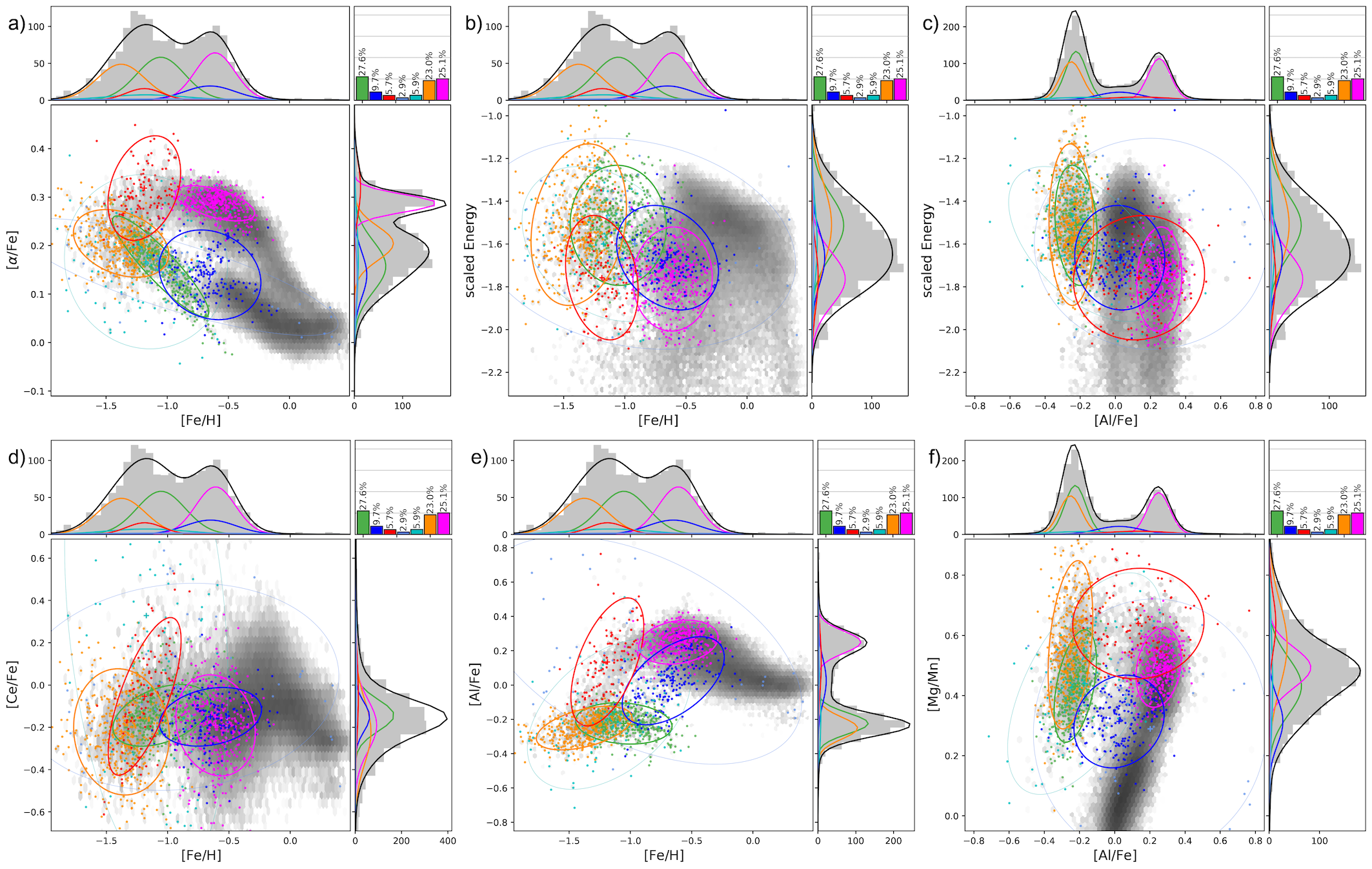}}
\end{center}
\caption{Distribution of Galaxy's eccentric populations in APOGEE-\textit{Gaia} sample in chemo-dynamical spaces. The GMM identified components are marked with $2\sigma$ ellipses. The \textit{Splash} and \textit{Aurora} are marked with magenta and red, The GS/E (GS/E$_1$ \& GS/E$_2$ enclosing the metal-richer and metal-poorer portion) components are marked with green and yellow. The \namesp is marked with blue. Two background components with low weight and no clear physical trend are marked with cyan and navy. The modelled distribution for each component and the total sum (black line) is marked in 1-dimensional histogram along each dimension at the top and the right side of each panel together with the background histogram marking the distribution of the stars used for the GMM. The weight of each component is marked at the top-right corner of each panel. 2-dimensional density histogram of the whole APOGEE-\textit{Gaia} dataset is shown as a greyscale background as a reference. Each panel shows different 2-dimensional projection of the original 6-dimensional chemo-dynamical space used for the GMM, \textit{a)}: [$\alpha$/Fe] vs [Fe/H], \textit{b)}: scaled energy vs [Fe/H], \textit{c)}: scaled energy vs [Al/Fe], \textit{d)}: [Ce/Fe] vs [Fe/H], \textit{e)}: [Al/Fe] vs [Fe/H], \textit{f)}: [Mg/Mn] vs [Al/Fe]. 
}
\label{fig:apogeeplanes}
\end{figure*}

\begin{figure*}
\begin{center}
{\includegraphics[width=185mm,angle=0,trim={ .cm .cm .cm 0cm},clip]{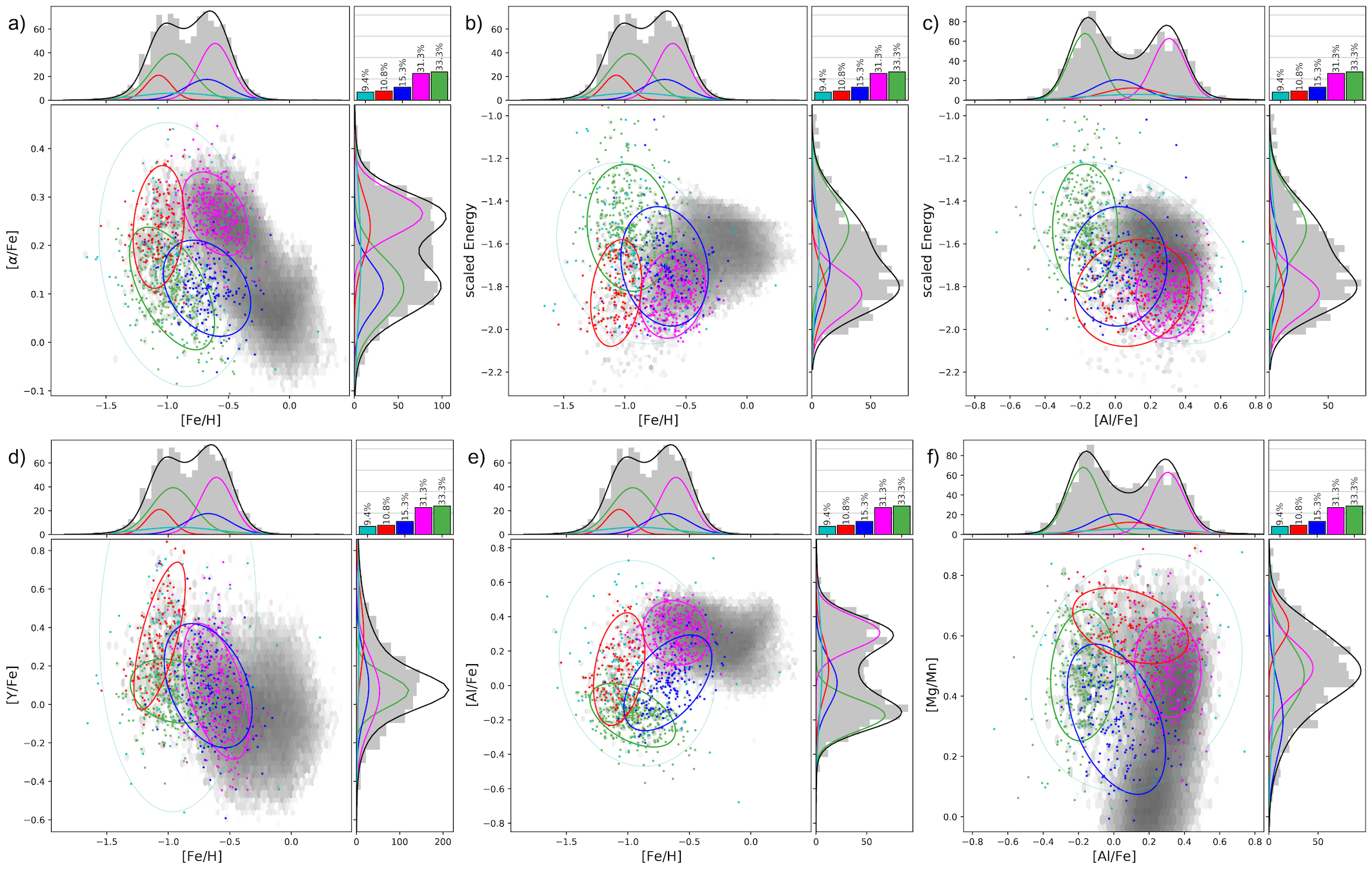}}
\end{center}
\caption{Distribution of Galaxy's eccentric populations in chemo-dynamical spaces similar to Figure~\ref{fig:apogeeplanes} but with GALAH-\textit{Gaia} sample. For GALAH-\textit{Gaia} sample 12-dimensional chemo-dynamical space is used for the GMM and the projections corresponding to Figure~\ref{fig:apogeeplanes} are shown. The GS/E component is marked with green (see Section~\ref{sec:results} for more detail). Panel \textit{d)} shows [Y/Fe] vs [Fe/H].  
}
\label{fig:galahplanes_1}
\end{figure*}

Figure~\ref{fig:aicbic} shows the scores from the BIC and AIC tests. For each model with $N$ Gaussians, the solid line gives the lowest score out of 100 XD runs. The dashed line is the median, and the dotted line is the highest score for each $N$ Gaussian case. Note that, although the BIC criterion provides a clear minimum, the AIC does not. The differences between AIC and BIC are discussed in standard texts on Bayesian model fitting using information criteria~\citep[e.g.,][]{Ge15}. Here, we merely note that model selection is a difficult task, without a clear-cut answer as to which information criteria works best.

\begin{figure*}
\begin{center}
{\includegraphics[width=185mm,angle=0,trim={ .cm .cm .cm 0cm},clip]{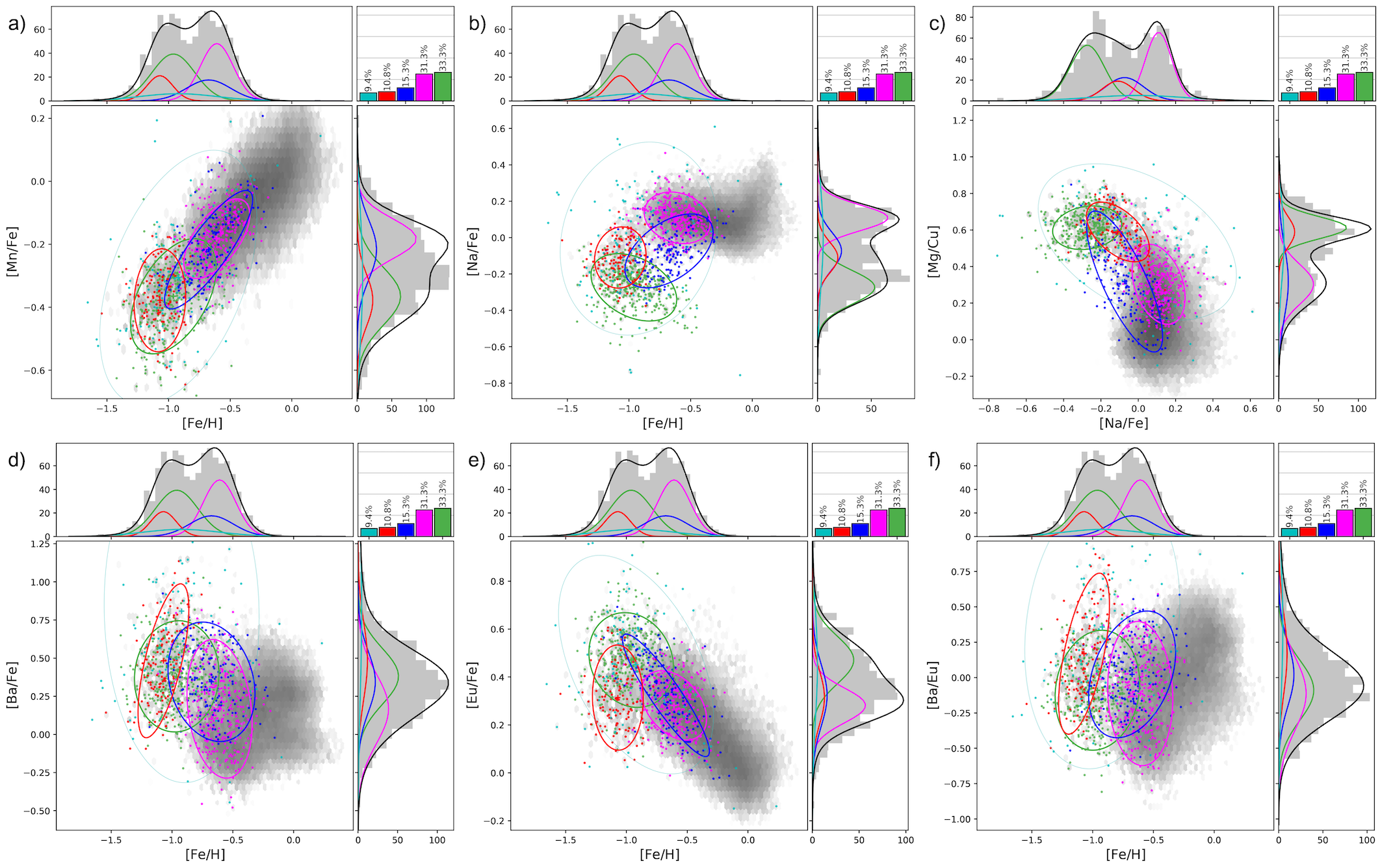}}
\end{center}
\caption{Distribution of Galaxy's eccentric populations in GALAH-\textit{Gaia} sample in chemo-dynamical spaces similar to Figure~\ref{fig:galahplanes_1} but for different 2-dimensional projection of the original 12-dimensional chemo-dynamical space used for the GMM, \textit{a)}: [Mn/Fe] vs [Fe/H], \textit{b)}: [Na/Fe] vs [Fe/H], \textit{c)}: [Mg/Cu] vs [Na/Fe], \textit{d)}: [Ba/Fe] vs [Fe/H], \textit{e)}: [Eu/Fe] vs [Fe/H], \textit{f)}: [Ba/Eu] vs [Fe/H].
}
\label{fig:galahplanes_2}
\end{figure*}

\begin{deluxetable*}{ccccccccc}
\tablecaption{Summary of GMM fit result for APOGEE-\textit{Gaia} sample} \label{tab:apogeefitresult}
\tablewidth{0pt}
\tablehead{
\colhead{Component} & \colhead{Weight} & \colhead{Count} & \colhead{[Fe/H]} & \colhead{[$\alpha$/Fe]} & \colhead{Energy} & \colhead{[Ce/Fe]} & \colhead{[Al/Fe]} & \colhead{[Mg/Mn]} \\
\colhead{} & \colhead{(\%)}  & \colhead{} & \colhead{} & \colhead{} & \colhead{($10^{5}$ km$^{2}$\,s$^{-2}$)} & \colhead{} & \colhead{} & \colhead{}
}
\startdata
\textit{GS/E$_1$} & 27.6 & 445 & $-1.05\pm0.20$ & $0.16\pm0.05$ & $-1.51\pm0.14$ & $-0.14\pm0.07$ & $-0.22\pm0.06$ & $0.43\pm0.09$ \\
{\it GS/E$_2$}\ & 23.0 & 396 & $-1.38\pm0.20$ & $0.20\pm0.03$ & $-1.51\pm0.19$ & $-0.22\pm0.15$ & $-0.25\pm0.06$ & $0.57\pm0.14$ \\
\textit{Splash} & 25.1 & 417 & $-0.61\pm0.16$ & $0.29\pm0.02$ & $-1.76\pm0.12$ & $-0.19\pm0.12$ & $0.25\pm0.06$ & $0.49\pm0.07$ \\
\textit{Aurora} & 5.7 & 112 & $-1.19\pm0.15$ & $0.32\pm0.05$ & $-1.76\pm0.15$ & $-0.05\pm0.19$ & $0.13\pm0.19$ & $0.64\pm0.09$ \\
\namesp & 9.7 & 177 & $-0.65\pm0.21$ & $0.14\pm0.05$ & $-1.66\pm0.12$ & $-0.15\pm0.07$ & $0.03\pm0.13$ & $0.31\pm0.08$ \\
back1 & 5.9 & 105 & $-1.17\pm0.33$ & $0.17\pm0.09$ & $-1.60\pm0.18$ & $0.33\pm0.63$ & $-0.17\pm0.22$ & $0.44\pm0.18$ \\
back2 & 2.9 & 54 & $-0.92\pm0.66$ & $0.14\pm0.06$ & $-1.60\pm0.25$ & $-0.01\pm0.25$ & $0.20\pm0.33$ & $0.29\pm0.22$ \\
\hline
\textit{GS/E$_{\mathrm{sum}}$} & 50.6 & 841 & $-1.20\pm0.25$ & $0.18\pm0.05$ & $-1.50\pm0.18$ & $-0.18\pm0.15$ & $-0.24\pm0.07$ & $0.49\pm0.14$ \\
back$_{\mathrm{sum}}$ & 8.8 & 159 & $-1.08\pm0.47$ & $0.16\pm0.08$ & $-1.59\pm0.22$ & $0.21\pm0.56$ & $-0.05\pm0.31$ & $0.39\pm0.21$ \\
\enddata
\tablecomments{ Weight is the GMM estimated weight of each component. Count is the number of stars assigned to each component based on the membership probability. See Section~\ref{sec:GMMview} for more detail. GS/E$_{\mathrm{sum}}$ row marks the combined properties of GS/E$_1$ and GS/E$_2$ (metal-richer and metal-poorer portion of the GS/E). back$_{\mathrm{sum}}$ row marks the combined properties of two background components, back1 and back2.}
\end{deluxetable*}

\begin{splitdeluxetable*}{ccccccccBccccccc}
\tablecaption{Summary of GMM fit result for GALAH-\textit{Gaia} sample} \label{tab:galahfitresult}
\tablewidth{0pt}
\tablehead{
\colhead{Component} & \colhead{Weight} & \colhead{Count} & \colhead{[Fe/H]} & \colhead{[$\alpha$/Fe]} & \colhead{Energy} & \colhead{[Ba/Fe]} & \colhead{[Y/Fe]} & \colhead{[Na/Fe]} & \colhead{[Al/Fe]} & \colhead{[Mn/Fe]} & \colhead{[Mg/Cu]} & \colhead{[Mg/Mn]} & \colhead{[Eu/Fe]} & \colhead{[Ba/Eu]} \\
\colhead{} & \colhead{(\%)}  & \colhead{} & \colhead{} & \colhead{} & \colhead{($10^{5}$ km$^{2}$\,s$^{-2}$)} & \colhead{} & \colhead{} & \colhead{} & \colhead{} & \colhead{} & \colhead{} & \colhead{} & \colhead{} & \colhead{}
}
\startdata
\textit{GS/E} & 33.3 & 354 & $-0.96\pm0.17$ & $0.11\pm0.06$ & $-1.52\pm0.15$ & $0.38\pm0.18$ & $0.07\pm0.08$ & $-0.27\pm0.09$ & $-0.17\pm0.09$ & $-0.37\pm0.09$ & $0.61\pm0.06$ & $0.47\pm0.11$ & $0.47\pm0.10$ & $-0.09\pm0.21$ \\
\textit{Splash} & 31.3 & 337 & $-0.61\pm0.13$ & $0.27\pm0.04$ & $-1.83\pm0.10$ & $0.17\pm0.23$ & $0.07\pm0.18$ & $0.11\pm0.07$ & $0.31\pm0.09$ & $-0.18\pm0.06$ & $0.30\pm0.11$ & $0.49\pm0.08$ & $0.28\pm0.07$ & $-0.11\pm0.25$ \\
\textit{Aurora} & 10.8 & 125 & $-1.07\pm0.10$ & $0.24\pm0.06$ & $-1.83\pm0.12$ & $0.48\pm0.25$ & $0.36\pm0.19$ & $-0.11\pm0.08$ & $0.10\pm0.16$ & $-0.38\pm0.08$ & $0.59\pm0.08$ & $0.63\pm0.06$ & $0.31\pm0.11$ & $0.17\pm0.28$ \\
\namesp & 15.3 & 168 & $-0.67\pm0.18$ & $0.11\pm0.05$ & $-1.71\pm0.14$ & $0.34\pm0.20$ & $0.10\pm0.16$ & $-0.07\pm0.10$ & $0.02\pm0.14$ & $-0.22\pm0.09$ & $0.32\pm0.19$ & $0.32\pm0.12$ & $0.32\pm0.13$ & $0.02\pm0.22$ \\
back & 9.4 & 106 & $-0.92\pm0.32$ & $0.18\pm0.14$ & $-1.64\pm0.21$ & $0.81\pm0.56$ & $0.43\pm0.49$ & $-0.01\pm0.27$ & $0.13\pm0.30$ & $-0.31\pm0.20$ & $0.53\pm0.21$ & $0.48\pm0.20$ & $0.45\pm0.23$ & $0.36\pm0.50$ \\
\enddata
\tablecomments{ Weight is the GMM estimated weight of each component. Count is the number of stars assigned to each component based on the membership probability. See Section~\ref{sec:GMMview} for more detail.}
\end{splitdeluxetable*}

\section{The Galaxy's eccentric constituents} \label{sec:results}

\subsection{GMM view of the local halo} \label{sec:GMMview}

GMM is an unsupervised probabilistic modelling technique. Naturally enough, it recovers the chemical signatures of the three known eccentric components of the Galaxy, namely the GS/E, the \textit{Splash} and \textit{Aurora}. This is in itself valuable as it confirms earlier results derived by extracting samples of these populations by hand-made cuts. Equally, this confirmation adds extra confidence to the GMM's results when the method breaks completely new ground.

For each GMM component in the APOGEE-\textit{Gaia} sample, the parameters describing the weight (fractional contribution), the mean and the standard deviation $\sigma$ of the model are presented in Table~\ref{tab:apogeefitresult}. The value for the combined GS/E population is also marked as GS/E$_{\mathrm{sum}}$. The sum of the two low-amplitude physically implausible components, associated with the background, (back$_{\mathrm{1}}$, back$_{\mathrm{2}}$) are also marked as back$_{\mathrm{sum}}$ for a reference. Figure~\ref{fig:apogeeplanes} shows several 2-dimensional projections of the original 6-dimensional chemo-dynamical APOGEE-\textit{Gaia} space used for the GMM. The best-fit model contains 7 Gaussian components shown with 2$\sigma$ ellipses. As a result of the fit, each star in the sample has a probability of belonging to an individual Gaussian component. By comparing these probabilities, the component membership of each star is established. There can be cases that a star has comparable probabilities between multiple components, making the classification less certain. Although this can result in slight deviation between the GMM estimated weight and the number of stars assigned to each component, we note that this deviation is within a small fraction. For our samples, there are $<1\%$ in APOGEE and $<2\%$ in GALAH that the difference between the highest and the second-highest component probabilities for each star is less than 0.05 (probability ranges from 0 to 1). 
While we use the stars with high orbital eccentricity ($e > 0.85$) for our GMM study, the density histogram of the whole APOGEE-\textit{Gaia} dataset is also shown as a 2-dimensional greyscale background for each projection to put the properties of the model Gaussian components into context.

We also apply the XD algorithm to GALAH-\textit{Gaia}, which include additional tracers. For the odd-Z and iron-peak elements, Na and Cu are included in addition to the abundances used in the fitting of the APOGEE-\textit{Gaia} sample. For the $s$-process, Y and Ba are used instead of Ce. In the case of the $r$-process, Eu is included. This is particularly desirable, as no $r$-process tracer is available in the APOGEE-\textit{Gaia} sample. Similar to the APOGEE-\textit{Gaia} sample, the parameters of each of the components identified by the GMM in GALAH are listed in Table~\ref{tab:galahfitresult}. Figures~\ref{fig:galahplanes_1} and \ref{fig:galahplanes_2} shows sets of 2-dimensional projections among the 12-dimensional GALAH sample space. 

The GMM for both samples finds strong evidence for 4 independent, physically plausible components. These comprise the three known constituents, together with an entirely new population called \name. Only one of these components is of accreted origin, namely the GS/E. This dominates the eccentric sample locally. Importantly, there is no clear chemical evidence for another major accretion event in the Solar neighborhood. This makes sense -- only the most massive accreted objects can fall deep into the potential well of the Galaxy. Innumerable minor accretion events have indeed occurred~\citep[e.g.,][]{Myeong2018clumps,Myeong2018_OmC,Koppelman2019,Ohare2020,Borsato2020,Limberg2021,Lovdal2022}, but their residues are -- relatively speaking -- inconsequential for the local sample.

There are other reassuring properties of the GMM results that add credibility to its decomposition. First, there is consistency between the results when applied to APOGEE-{\it Gaia} and GALAH-{\it Gaia} in terms of the number and properties of the components. Both datasets provide evidence for 4 physically-motivated components. Secondly, whenever there are elements in common between the two samples, then there is consistency in the results for the 4 components, as shown for example in Figs.~\ref{fig:apogeeplanes} and ~\ref{fig:galahplanes_1}. Thirdly, there is also consistency between APOGEE-{\it Gaia} and GALAH-{\it Gaia} for the same groups of elements. For example, in odd-Z ([Al/Fe], [Na/Fe]), $\alpha$- ([$\alpha$/Fe]) or iron-peak ([Mn/Fe]) elemental planes, the GMM clearly sees the same evolutionary trends in all 4 populations, irrespective of whether the APOGEE-{\it Gaia} and GALAH-{\it Gaia} datasets are used.

The only substantial difference is in the fractional contributions of the 4 components. For example, the GMM judges that GS/E comprises a majority of the local halo when applied to APOGEE-{\it Gaia}, but only a third in GALAH-{\it Gaia}. Here, we readily concede that the relative weightings of the populations are due to the surveys' distinct selection functions. For example, the  GALAH-{\it Gaia} sample is undernourished in stars below [Fe/H] $<-1.3$\,dex, so the GMM over-weights the {\it Splash}, the {\it Aurora} and the \name, while under-weighting the GS/E, as compared to APOGEE-{\it Gaia}.

Another interesting difference is that, in APOGEE-\textit{Gaia}, the GMM is powerful enough to split the GS/E into two separate, but closely-related, Gaussians. These correspond to the $\alpha$ plateau and the knee of the progenitor dwarf. In APOGEE-{\it Gaia}, the GS/E population is so well represented, especially at low metallicities, that the GMM can even break down the population into its {\it internal} chemical structure. 

Finally, in both the APOGEE-{\it Gaia} and GALAH-{\it Gaia}, the GMM finds low-weight \lq\lq background populations''. These are consistent with an outlier contribution, as they do not show physically plausible chemo-dynamical trends. For example, they have very large and thus physically implausible dispersion in most elements encompassing giant swaths of the abundance space. Since there will clearly be a discrepancy between the true distribution of each genuine Galactic population and the Gaussian model assumed by the GMM, residuals between the data and the model are expected. Such residuals across the sample space can easily be absorbed into a broad, low-weight and featureless background (Gaussian) density GMM component. 

We now discuss the GMM results for the 4 physically plausible components in turn. For the GS/E and the {\it Splash}, this is largely a confirmation and amplification of earlier results, though now obtained by an unsupervised learning method. For the {\it Aurora}, the GMM identifies completely new chemical trends, e.g. in Eu, Y and Ba, whilst \namesp is here identified and described for the very first time.

\subsection{GS/E}

The GS/E takes up $\approx51\%$ of eccentric stars in the APOGEE-\textit{Gaia} sample ($\approx33\%$ in the GALAH-\textit{Gaia} sample). GS/E has an \textit{ex situ} origin, having formed as a result of the disruption of a relatively massive galaxy. The mass of the progenitor dwarf has to be high to shrink and radialize the orbit~\citep{Vasiliev2022radialization}, so that the bulk of its stars are deposited into the inner regions of the Milky Way's halo. Only the most massive objects are able to sink sufficiently deep in the host's potential (e.g., dynamical friction being stronger for massive acquisitions) and contribute a sizeable fraction of their phase-mixed (and thus covering a significant volume) tidal debris in a given small region. Of all the components, the GS/E shows the highest mean orbital energy with the widest spread, as well as the lowest metallicity. This is also in accord with its \textit{ex situ} origin and with its debris being spread during the interaction (see e.g., panels b) and c) of Figures~\ref{fig:apogeeplanes} and \ref{fig:galahplanes_1}. 

With a progenitor mass of at least 1/10 of the Milky Way~\citep[e.g.,][]{Belokurov2018} at the time of merger, GS/E shows a metal-poor $\alpha$-knee at [Fe/H] $\approx -1.3$~\citep[e.g.,][]{Helmi2018,Mackereth2019}. Its inefficient star-formation is linked to its meagre enrichment in the odd-Z elements, which is reflected in the GS/E's low [Al/Fe] and [Na/Fe] level. 

Since our APOGEE-\textit{Gaia} sample extends down to lower metallicity, we can trace the GS/E's behavior at low metallicity better than in GALAH-\textit{Gaia}. In Figure~\ref{fig:apogeeplanes}, we see that the GS/E population splits into two. The two components have similar mean energy. However, the low metallicity portion (GS/E$_2$, yellow) shows a wider spread across the orbital energy range and also reaches down to lower orbital energies compared to its higher metallicity counterpart (GS/E$_1$, green). 
Note that even in the absence of dynamical friction, the dwarf progenitor's stars could cover a wide range of energies with leading and trailing debris occupying preferentially lower and higher energies correspondingly~\citep[see also,][for complications in the case of a massive and rotating progenitor]{Donghia2009}. This effect will blur any simple metallicity-energy correlation and could also result in the lower metallicity debris showing a wider spread, including the higher and lower ends of the orbital energy range.
GS/E$_1$ and GS/E$_2$ correspond to (i) the knee/decline in $\alpha$ and in [Al/Fe] with increasing contribution from SN Ia and (ii) high-$\alpha$ plateau and slowly rising [Al/Fe]. The two GS/E components also show different behaviour in Ni, another element sensitive to SN Ia explosions.

In the APOGEE-\textit{Gaia} data, the GS/E population covers a wide range of metallicities. So, we can expect its chemical distribution (e.g., [Fe/H], [$\alpha$/Fe], [Al/Fe], [Mg/Mn]) to be considerably non-Gaussian. Since GMM assumes its components are Gaussian, it struggles to reproduce such an extended component with a single Gaussian. This is why the GS/E population in our APOGEE-\textit{Gaia} sample splits into two separate but related Gaussians. For example, [Fe/H] $\approx -1.3$ is a pivotal point where GS/E's slope undergoes a small but noticeable change in [$\alpha$/Fe] and [Al/Fe] as a function of [Fe/H] (e.g., panel a) of Figure~\ref{fig:apogeeplanes}). GS/E's [Al/Fe] shows a mild increase along the metallicity range probed, reaching its peak around [Fe/H] $\approx -1.3$, and turns to a decreasing trend beyond this metallicity. This pattern is in good agreement with the earlier analysis of APOGEE data reported by e.g., \citet{Hasselquist2021} and \citet{Belokurov2022}. 

The unbiased GMM decomposition also recovers the fact that the GS/E is high in [Eu/Fe] and [Eu/Mg], consistent with previous studies~\citep{Aguado2021,Matsuno2021}. Eu and Mg are both good tracers for Core-Collapse Supernovae (CCSNe) contribution in star formation history as they are believed to be produced from massive stars most exclusively. But, in case of Eu ($r$-process element), Neutron Star Mergers (NSMs) are another considerable source of enrichment ~\citep[see e.g.,][]{Kobayashi2020}. These two $r$-process channels have distinct timescales -- CCSNe (NSMs) are known for prompt (delayed) $r$-process production, respectively. This means for a system with longer duration of star formation the [Eu/Mg] level will increase, as it will give more chance for the delayed Eu enrichment of NSMs to be reflected. Thus Eu trends mentioned above are a consequence of the longer star formation duration for the GS/E than the Milky Way. The Milky Way progenitor's has a much more vigorous and efficient star formation than GS/E progenitor.  

\subsection{Splash}

The \textit{Splash} is another major component in our high-eccentricity sample. Compared to GS/E, the orbital energy of the \textit{Splash} stars is also low. This is consistent with 
the idea that it is an \textit{in situ} population kicked out of the Galactic disk by the GS/E merger. 

The \textit{Splash} contributes $\approx25\%$ in APOGEE-\textit{Gaia} and $\approx31\%$ in GALAH-\textit{Gaia} sample. The differences between the two datasets in the fractional contribution of this population are likely due to the differences in the surveys' selection functions, particularly the metallicity range probed. For example, our GALAH-\textit{Gaia} sample has very few stars below [Fe/H]$<-1.3$\,dex. Since the GS/E population is dominant below this metallicity, GS/E's total weight in the GALAH sample appears to be lower compared to the APOGEE sample which extends down to lower metallicity. 

In comparison to GS/E, the \textit{Splash} population shows a distinctively high [Fe/H], [$\alpha$/Fe] and [Al/Fe] abundances which are the known characteristics of this  high-$\alpha$ \textit{in situ} population~\citep[][]{Nissen2010,Hawkins2015,Haywood2018,Dimatteo2019,Belokurov2020}. The observed spread in chemical element distributions are narrow in general for the \textit{Splash}, especially for [$\alpha$/Fe] and [Al/Fe], whilst the metallicity is 
limited to $-0.9<$[Fe/H]$<-0.3$\,dex. The main advance in this paper is here the chemical properties of the \textit{Splash} have been \lq\lq discovered'' by an unbiased GMM decomposition rather than extrapolated from hand-made cuts.

There is remarkable consistency across all chemical dimensions between the \textit{Splash} and the high-$\alpha$ disc. This is of course only to be expected -- the \textit{Splash} is born out of the evolved high-$\alpha$, \textit{in situ} component of the Galaxy.

\subsection{Aurora}

The local eccentric halo component with the lowest fractional contribution (marked in red, at $\approx6\%$ in APOGEE and $\approx11\%$ in GALAH) corresponds to the recently discovered \textit{Aurora} population~\citep[see][]{Belokurov2022}. This \textit{in situ} population is identified by the GMM primarily based on the chemical information, yet it shows behaviour entirely consistent with that described in \citet{Belokurov2022} where it was picked up based on kinematic evidence. For example, according to the GMM, in APOGEE DR17 \textit{Aurora} is limited to the metallicity range of $-1.5<$[Fe/H]$<-0.9$\,dex in line with the earlier study. \textit{Aurora} has one of the lowest mean orbital energy, similar to \textit{Splash}. However, given the apo-centre selection cuts imposed in this analysis, the main body of \textit{Aurora} likely has even lower energy and probably resides mostly in the inner Galaxy.

The unbiased GMM decomposition of both APOGEE and GALAH data offers a new, expanded chemical view of \textit{Aurora}, demonstrating clearly where and how this population stands out from the rest of the Milky Way. For example, at its high metallicity end, i.e. around [Fe/H]$\approx-1$, \textit{Aurora} exhibits the highest abundance ratios in $\alpha$-elements as well as [Al/Fe], [Ce/Fe], [Y/Fe], [Ba/Fe], [Ba/Eu] amongst all 4 local halo components. The GMM also detects strong correlations  between [Ce/Fe], [Al/Fe], [Y/Fe], [Ba/Fe] and [Fe/H]. Curiously, enhancement in the $s$-process elements has been seen in a number of currently existing GCs. For example, \citet{Meszaros2020} demonstrated the existence of $s$-process rich stars in a number of Galactic GCs, together with a strong correlation between Ce and metallicity. Similarly, \citet{Marino2009} reported bimodality in the $s$-process elements including Y and Ba, also accompanied by a strong correlation with metallicity, which they relate to the multiple cluster populations. These studies suggested that the $s$-process enhancement takes place in low- and intermediate-mass Asymptotic Giant Branch (AGB) stars. \citet{Belokurov2022} argued that \textit{Aurora}'s high dispersion and anomalous abundance in N, Al, O, Si could be linked to a significant contribution from massive star clusters (some of which may even survive to become Galactic GCs today). Here we provide additional evidence to support this hypothesis: the anomalous enrichment in odd-$z$ and $s$-process species together with metallicity correlations. 
 
\begin{figure*}
\begin{center}
{\includegraphics[width=185mm,angle=0,trim={ .cm .cm .cm 0cm},clip]{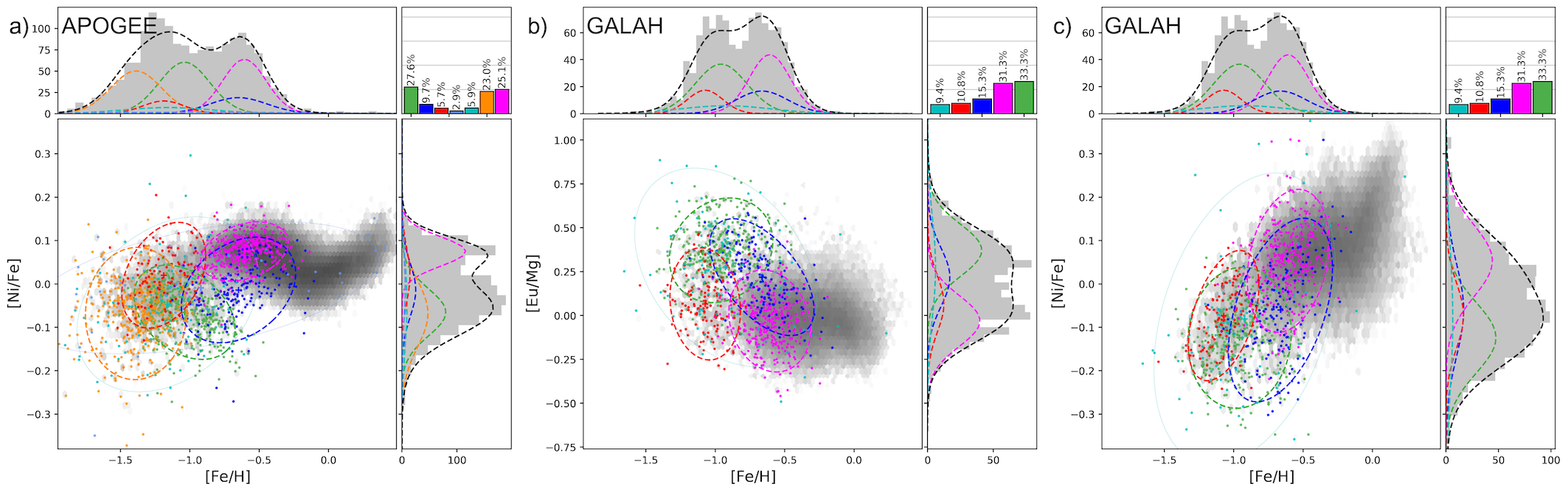}}
\end{center}
\caption{Distribution of GMM classified stars in APOGEE- \& GALAH-\textit{Gaia} sample in chemical dimensions ([Ni/Fe], [Eu/Mg]) that are not used in the GMM. Since [Ni/Fe] and [Eu/Mg] are not used in the GMM, the GMM result doesn't contain the modelled mean, covariance, and $\sigma$ for these dimensions. Instead, we use the GMM classified member stars for each component and fit a Gaussian representing their distribution in each projection. This fit is marked with dashed line. The color schema follows the Figure~\ref{fig:apogeeplanes} and \ref{fig:galahplanes_1}. \textit{a)}: [Ni/Fe] vs [Fe/H] for APOGEE-\textit{Gaia} sample, \textit{b)}: [Eu/Mg] vs [Fe/H] for GALAH-\textit{Gaia} sample, \textit{c)}: [Mg/Cu] vs [Na/Fe] for GALAH-\textit{Gaia} sample.
}
\label{fig:apgl_ni_eu_mg}
\end{figure*}

As far as the iron-peak elements are concerned, in particular those sensitive to SN Ia enrichment, as shown in Figure~\ref{fig:apgl_ni_eu_mg}, there is a positive [Ni/Fe]--[Fe/H] correlation amongst \textit{Aurora}'s stars. Interestingly this is a very different [Ni/Fe] trend compared to GS/E at a similar metallicity. Moreover a good number of \textit{Aurora} stars sit above GS/E in [Ni/Fe]. At the same time, [Mn/Fe] shows no apparent difference (see Figure~\ref{fig:galahplanes_2}). As \citet{Sanders2021} discuss, the plane of [Ni/Fe] and [Mn/Fe] provides a way to decipher SN Ia yields and their dependence on e.g. the host's star formation history. As Figure 9 of \citet{Sanders2021} shows, increased [Ni/Fe] at slowly varying metallicity and [Mn/Fe] may signal the growing contribution of massive SN Ia progenitors, which in turn may indicate a rise in star formation activity. Note that in massive Milky Way dwarfs [Ni/Fe] typically slowly decreases with metallicity, with the exception of the LMC where [Ni/Fe] shows an uptick connected to the most recent star formation burst~\citep[see][]{Hasselquist2021}.

Both [Eu/Fe] and [Eu/Mg] are markedly lower in  \textit{Aurora} compared to GS/E at the same metallicity (see Figures~\ref{fig:galahplanes_2} and ~\ref{fig:apgl_ni_eu_mg}). Recently, \citet{Naidu2022} noticed a similar difference between the proposed Heracles stars and GS/E. Using indirect evidence for star formation timescale in the two populations, they concluded that Heracles must have had a shorter (faster) SF activity compared to GS/E. \citet{Naidu2022} then explained the increase in [Eu/Mg] by the delay in the neutron star merger (NSM) contribution to Eu production, where the early phases of the enrichment in this $r$-process element are driven mostly by core collapse SN (CCSN). Perhaps both the [Ni/Fe] and the [Eu/Mg] trends in \textit{Aurora}'s stars are different pieces of evidence pointing in the direction of rapid SF activity in this ancient Milky Way population.

Comparing \textit{Aurora} to Heracles~\citep[c.f. this work and those by][]{Naidu2022, Horta2022}, it appears that in all chemical dimensions these two populations are close to indistinguishable. Yet, the proposed origins of the two are very different.  While \textit{Aurora} is created \textit{in situ}, Heracles is hypothesised to be an ancient merger event that occurred before the Milky Way-GS/E interaction. Heracles' stellar debris typically occupy high-eccentricity orbits (eccentricity $> 0.6$) but at a considerably lower energy range than the GS/E stars. \citet{Horta2021} used the apparent energy gap seen in the APOGEE dataset (see e.g., their Figure~4) to separate the GS/E and the Heracles stars. The Heracles' chemical signature is also different from that of the GS/E, for example, its [Mg/Fe] level remains consistently high while at the same range of metallicity, the GS/E shows a clear $\alpha$-knee, i.e. a decreasing trend with [Fe/H]. Additionally, there have been two separate studies proposing the existence of another major merger in our Galaxy, which supposedly took place, like Heracles, before the GS/E event. Based on the age-metallicity and the number of globular clusters residing in the very centre of the Milky Way, \citet{Kruijssen2019} argued for a major merger with a massive galaxy they named Kraken; a similar merger (Koala) was suggested by \citet{Forbes2020}. As these studies all suggest an older massive merger event (occurring before GS/E) with its debris mostly populating the inner Galaxy, a consensus is starting to emerge in the literature that Kraken, Koala and Heracles are in fact the same event~\citep[see e.g.,][]{Naidu2022}.

A recent study by \citet{Lane2022} (see e.g., their Section~6.3) pointed out that the dynamical characteristics of Heracles (e.g., the energy gap used to identify it as a distinct low-energy overdensity), is similar to the artificial density features induced by the APOGEE selection function. Indeed, no evidence for such an energy gap exists in the GALAH data which has a different (less bi-modal) selection function. Chemically, the low-[Fe/H], low-[Al/Fe] portion of \textit{Aurora} is remarkably similar to Heracles. Both have high [Mg/Fe], high [Mg/Mn], [Eu/Mg] $\approx0$~\citep[c.f. this study and][]{Naidu2022, Horta2022}. The only strong difference involves [Al/Fe]: \textit{Aurora} increases rapidly from sub-zero values (where it overlaps with Heracles) to extremely high [Al/Fe] $>0.5$\,dex, while Heracles stars have [Al/Fe] $<0$\,dex. Note however that the difference in [Al/Fe] between the two populations is artificial as the Heracles is selected to have negative [Al/Fe] to be consistent with an \textit{ex situ} origin~\citep[][]{Horta2021,Naidu2022, Horta2022}. A more in-depth detailed analysis can be carried out by comparing the \textit{Aurora} chemical properties as displayed in Figures~\ref{fig:apogeeplanes}, ~\ref{fig:galahplanes_1}, ~\ref{fig:galahplanes_2} and ~\ref{fig:apgl_ni_eu_mg} here to Heracles' chemical fingerprint as shown in \citet{Horta2022}. The similarities are striking. Despite being nearly identical chemically, Heracles and \textit{Aurora} appear to occupy very different regions of the orbital energy (and consequently Galactocentric radius) distribution. Heracles is selected to have low energy and reside in the bulge, while \textit{Aurora} has been identified - both here and in \citet{Belokurov2022} - outside of the bulge, amongst stars on high energy orbits. A comparison of the chemo-dynamical properties of \textit{Aurora} and Heracles seem to indicate that this is the same population that is not bound to either inner or outer Galaxy. Instead, probably, it extends beyond the Solar neighborhood but is mostly concentrated in the bulge, similar to the numerical examples shown in Figure~12 of \citet{Belokurov2022}.

\subsection{\name}

The GMM reconstruction of the local halo populations has revealed a new eccentric component that we dub \name. This previously unseen population occupies the metallicity range similar to that of \textit{Splash}, i.e. $-1<$[Fe/H]$<-0.3$\,dex. \name, which is shown in blue in the Figures in this paper, contributes $\approx10\%$ in APOGEE sample and $\approx15\%$ in GALAH sample\footnote{Lists of member stars are available electronically from the authors}. 
What sets \namesp apart is its relatively low $\alpha$-abundance level, it appears to be intermediate between the high-$\alpha$ sequence (in our sample represented by \textit{Splash}) and the GS/E at similar metallicity. Note also the difference in the orbital energy between \namesp and the high-$\alpha$ components (\textit{Aurora}, \textit{Splash}). \namesp stars have relatively higher orbital energy implying that it populates larger (outer) Galactic radii.

At its highest metallicity, across all elements \namesp connects seamlessly to the low-$\alpha$ (thin disk) sequence, yet unlike the disk it is highly eccentric. On the other hand, at its lowest metallicity \namesp resembles more GS/E than any other Galactic stellar populations. This is particularly obvious in [Al/Fe], [Ce/Fe], [Na/Fe], [Eu/Fe], [Ni/Fe], [Eu/Fe] and [Eu/Mg]. Could \namesp be a new accreted component of the halo? Judging by its high overall [Al/Fe], probably not. Its [Al/Fe] levels are higher than GS/E's but not as high as the high-$\alpha$ population. Moreover, \namesp exhibits a clear correlation between [Al/Fe] and [Fe/H] with a less steep slope compared to that of \textit{Aurora}. The [Al/Fe] signature of \namesp are replicated in the other odd-$z$ element available as part of this analysis, Na. Based on the elevated [Al/Fe] and [Na/Fe] values as well as their correlation with [Fe/H] we follow the ideas of \citet{Nissen2010} and \citet{Hawkins2015} and conclude that \namesp was created \textit{in situ}. As shown by \citet{Hasselquist2021} no Milky Way dwarf no matter how massive can reach as high levels of [Al/Fe] attained by the \name.

Another interesting trend in \namesp population is a possible bifurcation in [$\alpha$/Fe]--[Fe/H] plane visible in APOGEE sample (Figure~\ref{fig:apogeeplanes}, panel a) at around [Fe/H]$\approx-0.5$. A subset of \namesp stars appear to form a branch between the higher-[$\alpha$/Fe] with higher-[Fe/H] and the lower-[$\alpha$/Fe] with lower-[Fe/H] region in the [$\alpha$/Fe]--[Fe/H] plane. These stars also show relatively higher [Al/Fe] abundances than the rest of \namesp stars with comparable metallicity.
This feature could be in line with the ``Bridge'' region suggested by \citet{Ciuca2021} as a transition region connecting the old thick (high-$\alpha$) and young thin (low-$\alpha$) disks, with a possible age gradient (older at higher [$\alpha$/Fe], younger at lower [$\alpha$/Fe]). This Bridge region doesn't exactly overlap with \name's possible higher-$\alpha$ subbranch~\citep[Bridge region covers slightly higher metallicity and spans a very broad range of metallicity; see e.g., Figure 6 of][]{Ciuca2021}. Yet, the general idea of the transition signature running from higher-[$\alpha$/Fe] with higher-[Fe/H] to lower-[$\alpha$/Fe] with lower-[Fe/H] could be in line with \name's possible higher-[$\alpha$/Fe] subbranch feature, as a trace of gradual gas mixing (polluting) between the leftover gas from high-$\alpha$ (think) disk and the gas from GS/E merger (more discussion on \name's origin in Section~\ref{sec:eosorigin}). 
We should also consider the possible blending from GMM fitting between \namesp and the high-$\alpha$ disk populations, as these higher-[$\alpha$/Fe] \namesp stars are relatively closer to the high-$\alpha$ disk population in the chemical space.

\name's low-$\alpha$ abundance combined with its intermediate odd-Z abundance level also places \namesp stars in a unique region of [Mg/Mn]--[Al/Fe] and [Mg/Cu]--[Na/Fe] planes (panel f) of Figures~\ref{fig:apogeeplanes} and \ref{fig:galahplanes_1}, panel c) of Figure~\ref{fig:galahplanes_2}). These planes have been successfully used to discern the less-evolved (and/or \textit{ex situ}) and more-evolved (\textit{in situ}) stellar populations. As a less evolved portion of the low-$\alpha$ \textit{in situ} population, \namesp appears between the region occupied by GS/E and the bulk of the low-$\alpha$ \textit{in situ} stars (shown as a 2D density histogram in the Figures). \namesp also sits below the region where \textit{Aurora} is located in this plane emphasising a clear difference in their $\alpha$ abundance levels. Although both \namesp and \textit{Aurora} are likely to be the Galaxy's \textit{in situ} components, clear differences between them in a number of key elemental trends such as  metallicity, $\alpha$, and odd-Z suggests different formation channels for \namesp and \textit{Aurora}. In several of the chemical projections \namesp also displays a correlation with metallicity, i.e. [Mn/Fe], [Eu/Fe], [Ni/Fe], [Mg/Cu] all show trends with [Fe/H].

The chemical trajectory of \namesp is unique, its chemical evolution appears to have started from the gas polluted by the GS/E and evolved into a typical low-alpha (thin disk) population.

\section{Discussion and Summary} \label{sec:summ}

\subsection{Clues to \name's origin} \label{sec:eosorigin}

Deciphering the \name's origin may help us to solve the conundrum of the (trans)formation of the Milky Way's disc.
A part of the puzzle is the existence of a striking chemical bi-modality: two nearly parallel but well separated sequences are detected in the disc’s $\alpha$-[Fe/H] plane at a wide range of stellar metallicities and Galactocentric distances~\citep[e.g.][]{Fuhrmann1998, Feltzing2003,Adibekyan2013,Bensby2014,Hayden2015}. The high-$\alpha$ sequence and the low-$\alpha$ sequence have been linked to the so-called ``thick'' and ``thin'' discs respectively, although given the recent structural analysis it may be more appropriate to label the discs “inner” and “outer”~\citep[see e.g.][]{Haywood2013,Rix2013}. Chronologically, the high-$\alpha$ and the low-$\alpha$ discs appear distinct: the former built up most of its mass before 8-10 Gyr ago, the epoch around which the first cinders of star-formation have been detected in the thin disc which has been growing steadily ever since~\citep[see e.g.][]{Haywood2018phylogeny,Fantin2019}.  

Several scenarios have been proposed to explain the $\alpha$-dichotomy in the Galactic disc. For example, the “two-infall” model requires that in the distant past an episode of gas accretion diluted the Milky Way’s reservoir, resetting its metallicity so that the subsequent star-formation could proceed along the lower $\alpha$-sequence starting from a lowered value of [Fe/H]~\citep[see][]{Chiappini1997,Chiappini2009,Grisoni2017}. Alternatively, the observed $\alpha$-bimodality can be produced via radial migration and “churning” of the stars created along the two $\alpha$-sequences in distinct (broadly speaking, inner and outer) parts of the disc~\citep[see][]{Sellwood2002,Roskar2008,Schonrich2009,Minchev2013}. A new idea was proposed recently where the two spatially overlapping $\alpha$-sequences can start forming contemporaneously: massive dense gas clumps dominate the production of stars with high $\alpha$-abundance, while the smooth distributed star-formation is responsible for the low $\alpha$-abundance~\citep[][]{Clarke2019}. Finally, it is now also possible to form galaxies with $\alpha$-bimodal discs in Cosmological hydrodynamical simulations. Curiously, in the simulated discs, the $\alpha$-dichotomy is generally linked to ancient massive gas-rich accretion events~\citep[see e.g.][]{Brook2012,Grand2018,Mackereth2018,Buck2020}, however the details might vary. For example, in \citet{Agertz2021} around redshift z=1.5 a gas-rich accretion creates an outer low-$\alpha$ disc misaligned with respect to the earlier-formed inner high-$\alpha$ one; after a while, the two discs line up, leading to spatially overlapping low- and high-$\alpha$ stellar populations~\citep[see][]{Renaud2021_VintergatanIII}. In \citet{Grand2020}, an analogue of the thick disc exists prior to the GS/E event but this violent interaction contributes to its additional heating; moreover, the gas-rich GS/E merger induces a star-burst in the central MW which adds between 20\% and 60\% stellar mass to the thick disc. As hinted by these recent studies based on cosmological hydrodynamical simulations, the Galactic discs and the GS/E event are linked in terms of their kinematics, chemical composition, and chronology.

Curiously, both \citet{Renaud2021_VintergatanIII} and \citet{Grand2020} discuss the emergence of an {\it in situ} halo-like population distinct from \textit{Splash} and not much dissimilar to \namesp around the time of the GS/E accretion. In \citet{Grand2020}, a GS/E-like merger both splashes the pre-existing disk and induces a starburst in the centre of the Galaxy which leads to creation of a puffed-up halo-like \textit{in situ} component. The simulated \textit{Splash} would be expected to have a wide range of ages truncated around the time of the GS/E arrival, while the starburst population's SFH would peak at the epoch of the interaction. In \citet{Renaud2021_VintergatanIII}, the GS/E-like merger not only triggers star formation in the mis-aligned outer gaseous disk but also kick-starts its  re-orienting with the pre-existing inner one. The two gas disks align rapidly but the first stars that were born in the outer disk are left behind in a strongly tilted configuration. The early outer disk stars subsequently phase-mix, and at the present day lack any appreciable angular momentum -- the memory of their particular initial condition. The scenario of \citet{Renaud2021_VintergatanIII} is remarkable in that it predicts the existence of an \textit{in situ} halo-like population with a narrow age distribution and the chemistry similar to the metal-poor low-$\alpha$ disk.

In the two-infall scenario, a variation of which is played out in the simulations discussed above, the re-setting of the disk's chemical sequence is facilitated by the accretion of fresh, poorly-enriched gas from the medium around the host. It is also possible to directly invoke the accretion and recycling of the gas from the GS/E progenitor~\citep[see][]{Grand2020,Palla2020}. The dwarf is not required to donate its gas (which may not be enough to make a huge difference), it merely needs to enrich the gas surrounding the host. The exact amount of the heavy elements the dwarf can infuse into the condensing circumgalactic medium (CGM) will depend on the state of the gas and the parameters of the host-satellite interaction. Importantly, however, the mass loading factor of the pre-enriched outflows are predicted to be higher in the dwarf galaxy compared to the more massive host~\citep[][]{Mitchell2020,Pandya2021}. This is because the energetics at the feedback site do not depend on the host's global properties, but the restoring force (gravity) does. This idea may have some support in the chemical properties of \name. As Figures~\ref{fig:apogeeplanes},~\ref{fig:galahplanes_1} and~\ref{fig:galahplanes_2} demonstrate, at its lowest metallicity, \namesp is closest to the metal-rich portion of the GS/E sequence.

\subsection{Caveats}

Our study uses GMM to analyse unlabelled data in hyper-dimensional space consisting primarily of chemical information. Our modelling is based on the assumption that any sub-population found in the data follows a Gaussian distribution in the data space of choice. Unfortunately there is no guarantee that this is actually true. If a real physical trend in the data deviates severely from Gaussian distribution (e.g., a stellar population in [$\alpha$/Fe]--[Fe/H] with changing $\alpha$ abundance slope), this single stellar population can be segmented into multiple Gaussian components in the GMM fit. 
Any hard boundary cut applied to the data can also degrade the GMM performance. Therefore, we avoid the use of any dimension in our GMM if this dimension is directly used for any selection cuts adopted. 

Populations seen in our sample are the high-eccentricity tails of their actual orbital distribution. For each population, this will bias the weight of the population assigned by the GMM depending on the degree of radial anisotropy. For example, \textit{Aurora} is shown to be isotropic with its velocity dispersion, i.e., $\sigma_{\mathrm{R}} \approx \sigma_{\mathrm{z}} \approx \sigma_{\mathrm{\phi}}$~\citep[see][for more detail]{Belokurov2022}. In contrast, GS/E or \textit{Splash} are known for their strong radial anisotropy~\citep[e.g.,][]{Belokurov2018,Myeong2018_actionHalo,Myeong2018_SauGC,Belokurov2020}. Such a difference is reflected in their orbital eccentricity distribution, and different proportions of each population will be included in our high-eccentricity sample. While we have avoided making a cut on energy, the apocenter $> 5$\,kpc cut puts a slightly softer lower limit on the orbital energy range in our sample. This can bias the components' properties if they are found near this lower energy boundary. For example, \textit{Aurora} and \textit{Splash} appear to extend down to the lowest energies, close to the limit imposed.

The reliability of chemical measurements varies with all of the atmospheric parameters and generally deteriorates towards the lower end of the metallicity range. To minimize the effect of varying uncertainty, we adopted the XD algorithm for our GMM which can take account of measurement errors associated with the observed data. However, of course, the systematic (unreported) errors will still affect the results of our modelling. The metallicity probed by both APOGEE and GALAH is truncated earlier as both surveys struggle in the low-[Fe/H] regime. The metallicity limit can also confuse the recovery of the true stellar populations in the data, especially for those components that extend far down in [Fe/H], e.g. GS/E and \textit{Aurora}.

\subsection{Summary}

Our study provides one of the first attempt to assemble an unbiased census of the main halo components in the vicinity of the Sun. To carry out this experiment, we use the Extreme Deconvolution implementation of the Gaussian Mixture Model. To help the GMM concentrate on the stellar populations of interest, we apply an eccentricity cut thus limiting the sample to the halo stars only. The GMM model provides results that show consistency between APOGEE and GALAH in terms of the number and the properties of the components recovered. Moreover, when comparing the detailed properties of the individual components in APOGEE and GALAH we see clear consistency for the same chemical elements and for the same groups of elements (e.g. the [Al/Fe] behaviour is consistent with the [Na/Fe] behavior). Small differences in fractional contributions of the components between APOGEE and GALAH are likely due to the surveys' distinct selection functions.

The reliability of the GMM decomposition reported here can be established by comparing the chemical properties of the known and well-studied halo components to those reported in the literature. For example, our view of the GS/E chemical fingerprint is in full agreement with the recent studies~\citep[see e.g.][]{Das2020,Hasselquist2021,Buder2021,Aguado2021,Matsuno2021,Naidu2022,Horta2022,Carrillo2022}. Similarly, we demonstrate that across all chemical dimensions investigated, \textit{Splash} is indistinguishable from the high-$\alpha$ (thick disk) of the Milky Way. This is of course in agreement with the proposed origin of this \textit{in situ} halo population~\citep[see e.g][]{Bonaca2017,Haywood2018,Gallart2019,Dimatteo2019,Belokurov2020}.

While it is reassuring to see agreement between the earlier attempts to decipher the chemical behavior of the halo denizens and our unsupervised analysis, the results of this work are not limited to mere confirmation. Let us summarise the most interesting new findings here. 

\begin{itemize}

    \item The GMM finds only four independent halo components dominating the Solar neighborhood, three of these are previously known, i.e. \textit{Aurora}, \textit{Splash} and GS/E, and one is new. There is strong evidence that the newly discovered population \namesp is of \textit{in situ} origin. Therefore, in the Solar neighbourhood, our analysis detects no evidence for tidal debris from any additional large accretion event.

    \item As reported by the GMM, the multi-dimensional chemical view of the ancient \textit{in situ} component \textit{Aurora}~\citep[see][]{Belokurov2022} agrees remarkably well with the chemical properties of the alleged merger remnant known as Heracles~\citep[][]{Horta2021,Naidu2022,Horta2022}. The only noticeable difference appears to stem from the fact that in the literature the Heracles stars are identified using a hard-cut in [Al/Fe] while here we dispense with selections in the chemical space. Our study preserves the integrity of the stellar populations in the elemental abundance space and demonstrates that in [Al/Fe]--[Fe/H] dimension \textit{Aurora} evolves rapidly from low [Al/Fe] to anomalously high [Al/Fe]. This behaviour is predicted for a rapidly star-forming self-enriching galaxy with a mass similar to that of the Milky Way~\citep[see][]{Horta2021} and thus obviates the necessity for introducing an additional merger event to explain this stellar population.

    \item The chemical properties of the discovered \textit{in situ} born structure \namesp evolve significantly as a function of metallicity. At its lowest [Fe/H] \namesp does not resemble any of the known \textit{in situ} populations, especially those that are predicted to have formed early, i.e. \textit{Aurora} or \textit{Splash}. Instead, the low-[Fe/H] \name's stars look similar to the most metal-rich GS/E members. However, at its highest [Fe/H] \namesp connects seamlessly to the chemical sequence  of the low-$\alpha$ (thin disk) stars. This strong evolution is exhibited as clear correlation between e.g. [Al/Fe], [Mn/Fe], [Eu/Fe], [Ni/Fe] and [Fe/H]. We speculate that \namesp appears to originate from the gas polluted by the GS/E and evolved to resemble the (outer) thin disk of the Milky Way.

\end{itemize}

The unsupervised view of the local halo the GMM provides also allows us to decipher many new and fine details of the four components detected. For example, in APOGEE, the GS/E is split into two components that primarily differ in metallicity but trace distinct epochs in the life of the progenitor galaxy, i.e. one corresponding to the high-$\alpha$ plateau and slowly rising [Al/Fe] and another to the knee/decline in $\alpha$ and in [Al/Fe] due to the increased SN Ia contribution. We also see that the two GS/E components behave differently in Ni (iron-peak element particularly sensitive to variations in SN Ia yields). Curiously, the GS/E trends in [Ni/Fe]--[Fe/H] space are distinct from that of \textit{Aurora} at similar metallicity. In both APOGEE and GALAH, \textit{Aurora} attains the highest abundance ratios in $\alpha$, [Al/Fe], [Ce/Fe], [Y/Fe], [Ba/Fe], [Ba/Eu] of all 4 components. Moreover, it shows strong correlations between [Ce/Fe], [Al/Fe], [Y/Fe], [Ba/Fe] and [Fe/H]. We interpret such behaviour as additional evidence that massive star clusters may have contributed significantly to the chemical evolution of \textit{Aurora}~\citep[as argued in][]{Belokurov2022}. We note that \textit{Aurora} has significantly lower [Eu/Fe] and [Eu/Mg] compared to GS/E at similar metallicity, the trend that has been claimed previously for Heracles by \citet{Naidu2021}.

Deciphering the origin of \namesp with the help of numerical simulations, both those currently available such as FIRE~\citep{Hopkins2014}, Auriga~\citep{Grand2017Auriga}, VINTERGATAN~\citep{Agertz2021} and ARTEMIS~\citep{Font2020ARTEMIS}, as well as those upcoming, is the obvious next step. Currently the feedback physics remains somewhat unconstrained but there may be routes to test the proposed formation mechanism of \name, in particular, the role of the GS/E progenitor galaxy in the pollution of the circum-host gas and the triggering of the \textit{in situ} star formation. The (relative) stellar ages~\citep[see e.g.][]{Conroy2022,Xiang2022} also provide valuable information and will help to expand the current analysis to a chrono-chemo-dynamical study. Undoubtedly, detailed information from the upcoming large spectroscopic surveys (e.g., DESI, 4MOST, WEAVE, SDSS-V) and astrometric surveys (e.g., \textit{Gaia}, LSST) will open up new arenas for decoding the genealogy of the Galaxy.

\begin{acknowledgments}
We thank the anonymous reviewer for valuable comments and suggestions which helped to revise/improve this manuscript.
We thank Andrey Kravtosv, Oscar Agertz, Florent Renaud, Martin Rey, 
Yeisson Osorio, Madeline Lucey, Julianne Dalcanton, Melissa Ness, Keith Hawkins, 
Adrian Price-Whelan, Henry Leung, Lucy Lu, Jason Sanders, Harley Katz, Giuliano Iorio, Chervin Laporte, and Iulia Simion for enlightening discussions that 
helped to improve this manuscript.
GCM acknowledges the support from the Harvard-Smithsonian Center for Astrophysics 
through the CfA Fellowship. GCM and DA thank the Center for Computational Astrophysics 
at Flatiron Institute for the warm hospitality during their visit.
DA acknowledges support from ERC Starting Grant NEFERTITI H2020/80824.

This research made use of data from the European Space Agency mission Gaia
(\url{http://www.cosmos.esa.int/gaia}), processed by the Gaia Data
Processing and Analysis Consortium (DPAC,
\url{http://www.cosmos.esa.int/web/gaia/dpac/consortium}). Funding for the
DPAC has been provided by national institutions, in particular the
institutions participating in the Gaia Multilateral Agreement. 

This paper made used of the Whole Sky Database (wsdb) created by Sergey Koposov 
and maintained at the Institute of Astronomy, Cambridge with
financial support from the Science \& Technology Facilities Council
(STFC) and the European Research Council (ERC).
\end{acknowledgments}

%

\vspace{5mm}


\software{AGAMA~\citep{Vasiliev_AGAMA2019},
          astropy~\citep{2013A&A...558A..33A,2018AJ....156..123A},
          Extreme Deconvolution~\citep{Bovy2011_XD},
          galpy~\citep{Bovy2015_galpy},
          scikit-learn~\citep{scikit-learn}
          }







\bibliography{MW_eccentric_constituents}{}
\bibliographystyle{aasjournal}



\end{document}